\documentclass[aps,prb,twocolumn,superscriptaddress]{revtex4-1}
\usepackage{amsmath, graphicx, subcaption,dcolumn,bm}
\DeclareMathOperator{\Tr}{Tr}
\usepackage{hyperref}
\usepackage[normalem]{ulem}
\usepackage[dvipsnames]{xcolor}
\usepackage{soul}
\definecolor{purp}{HTML}{832591}
\DeclareCaptionJustification{justified}{\leftskip=0pt \rightskip=0pt \parfillskip=0pt plus 1fil}
\hypersetup{
  colorlinks,
  allcolors = {purp},
  citecolor = {purp},
  allbordercolors = {white},
}
\begin{document}

\title{Experimental Measures of Topological Sector Fluctuations in the F-Model}

\author{Daan M. Arroo}
\affiliation{London Centre for Nanotechnology and Department of Physics and Astronomy, University College London, 17-19 Gordon Street, London WC1H 0AH, United Kingdom}

\author{Steven T. Bramwell}
\affiliation{London Centre for Nanotechnology and Department of Physics and Astronomy, University College London, 17-19 Gordon Street, London WC1H 0AH, United Kingdom}

\begin{abstract}

The two dimensional F-model is an ice-rule obeying model, with a low temperature antiferroelectric state and high temperature critical Coulomb phase. Polarization in the system is associated with topological defects in the form of system-spanning windings which makes it an ideal system on which to observe topological sector fluctuations, as have been discussed in the context of spin ice and Berezinskii-Kosterlitz-Thouless (BKT) systems. In particular, the F-model offers a useful counterpoint to the BKT transition in that winding defects are energetically suppressed in the ordered state, rather than dynamically suppressed in the critical phase. In this paper we develop Lieb and Baxter's historic solutions of the F-model to exactly calculate relevant properties, several apparently for the first time. We further calculate properties not amenable to exact solution by an approximate cavity method and by referring to established scaling results. Of particular relevance to topological sector fluctuations are the exact results for the applied field polarization and the ‘energetic susceptibility’. The latter is a both a measure of topological sector fluctuations and, surprisingly, in this case, a measure of the order parameter correlation exponent. In the high temperature phase, the temperature tunes the density of topological defects and algebraic correlations, with the energetic susceptibility undergoing a jump to zero at the antiferroelectric ordering temperature, analogous to the ‘universal jump’ in BKT systems. We discuss how these results are relevant to experimental systems, including to spin ice thin films, and, unexpectedly, to three-dimensional dipolar spin ice and water ice, where we find that an analogous ‘universal jump’ has previously been established in numerical studies. This unexpected result suggests a universal limit on the stability of perturbed Coulomb phases that is independent of dimension and of the order of the transition. Experimental results on water ice Ih are not inconsistent with this proposition. We complete the paper by relating our new results to experimental studies of artificial spin ice arrays.

\end{abstract}


\maketitle

\section{Introduction}

The two-dimensional F-model is an idealized model of an antiferroelectric that was introduced by Rys~\cite{Rys} in 1963. Exact solutions by Lieb~\cite{Lieb} and Sutherland~\cite{Sutherland1967} in 1967 and Baxter~\cite{Baxter1973} in 1973 later placed it the small class of exactly solved models in statistical mechanics~\cite{Baxter}. Theoretical interest in the F-model has long been sustained by its relevance to surface roughening (BCSOS model~\cite{vanBeijeren}), to quantum many body problems~\cite{Nijs}, and as an exemplar of an infinite order transition~\cite{Weigel, Keesman2016,Keesman}. 

\begin{figure}
	\centering
	\includegraphics[width=1.0\linewidth]{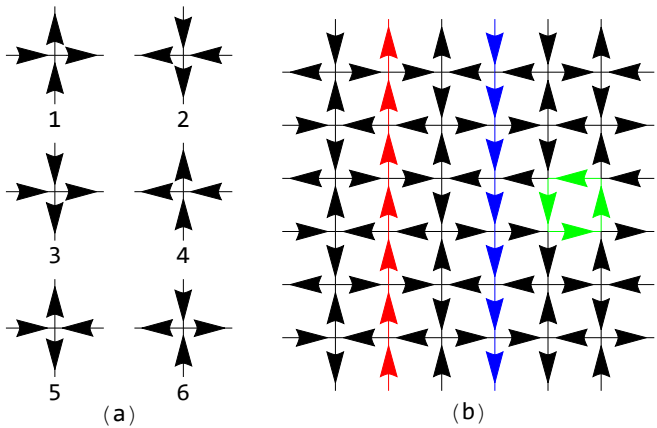}
	\caption{(a) The six vertices of the F-model, in which the vertices with no net polarization (5 and 6) are lower in energy than the remaining vertices (1, 2, 3, 4) by an energy gap $\epsilon$. (b) An excited F-model configuration with no net polarization. Excitations may consist of closed loops (green) which carry no net polarization as well as system-spanning loops (red and blue), which ``wrap'' the periodic boundary conditions and carry a net polarization.}
	\label{fig:vertexfiglabels}
\end{figure}

The F-model is an energetic ice-type model where different vertices may be assigned different energies (Fig. \ref{fig:vertexfiglabels}). In the F-model, out of the six ice-rule obeying vertices, the two that carry no net polarization lie lowest in energy by an amount $\epsilon$ and these form an antiferroelectric ground state below $T_{\rm 0} = \epsilon/\ln 2$. The high temperature phase is a critical six-vertex state, or Coulomb phase, analogous~\cite{Knops} to the low temperature critical state of a Berezinskii-Kosterlitz-Thouless (BKT) system~\cite{B,KT}. With inversion of the temperature scale, the F-model and BKT transitions share several, but not all, critical properties~\cite{Weigel}.

In common with the low temperature  state of ice, the high temperature Coulomb phase of the F-model has topological degeneracy. The F-model further has the unusual property that it has no ice rule defects and its polarization at all temperatures is entirely controlled by the excitation of system-spanning windings, which are global topological defects (see Fig. \ref{fig:vertexfiglabels}). This makes it an ideal system by which to study topological sector fluctuations, as defined for spin ice~\cite{Macdonald, Jaubert} and the two dimensional Coulomb gas at the BKT transition~\cite{Faulkner}. There the topological sectors are associated with harmonic winding modes of the magnetization and electric field respectively. The analogy with spin ice is a direct one, yet the F-model has the attractive quality that it is free from the complication of ice-rule defects that compete with the topological sector fluctuations in spin ice at finite temperatures. The analogy with BKT physics is less direct. The BKT transition in the 2D Coulomb gas is an ergodicity breaking -- a change in the phase space explored by a system with local dynamics -- where charge confinement suppresses the topological sector fluctuations~\cite{Faulkner}. The F-model offers a counterpoint to this, in that its system-spanning polarization windings are energetically suppressed in the low temperature (ordered) state, rather than dynamically suppressed in the critical state. It should be emphasised that because the energy of system-spanning windings scale as $O(L)$ (where $L =\sqrt{N}$ is the system size for a system of $N$ vertices), they are completely suppressed below $T_0$ in the thermodynamic limit, such that the ordered state of the F-model has zero direct susceptibility (in contrast, the staggered susceptibility of the order parameter remains finite). 

In part II of this paper we determine experimentally relevant properties of the F-model by developing the exact solution, as given by Baxter~\cite{Baxter1973, Baxter}. Most surprisingly, the exact field and temperature-dependent polarization has never previously been calculated explicitly (to our knowledge), while the exact specific heat has been exhibited a handful of times, but only in zero field~\cite{Wu,Weigel,Bovo2}. We particularly focus on the `energetic susceptibility'~\cite{Fisher1962} (derived from the direct susceptibility), which, in the analogous case of spin ice, has been shown to be a sensitive measure of topological sector fluctuations~\cite{Macdonald, Jaubert}.  We find that the energetic susceptibility has a further very interesting and surprising property in the F-model, being proportional to the correlation exponent $\eta(T)$ of the staggered order parameter. To complement the exact results we further develop the cavity model method of Foini and colleagues \cite{Foini2013},  to approximately calculate the order parameter susceptibility, which is not amenable to exact analysis.

In part III of the paper we turn our attention to experimental applications of the present results, focusing on the case of ice-type systems. In general, the application of highly idealized spin or vertex models to experimental systems is justified  by renormalization group theory. Starting with the idealised Hamiltonian, one can imagine adding small perturbations to make it more realistic. If these are relevant perturbations, then they will typically change the universality class near to a fixed point, but leave it unchanged elsewhere, such that the idealized model describes experiment over much of the phase diagram. 

In the particular case of the F-model applied to ice rule systems, the most important perturbation to consider is the thermal excitation of ice-rule breaking defects. These will indeed change the universality class of the transition, typically rendering it second order. However, in many such systems the defect density is extremely small in the relevant temperature range. For example, it is parts per billion in proton disordered ice Ih, strictly negligible in proton ordered ice-II~\cite{Whalley, Shepard}, and parts per thousand in spin ice~\cite{Kaiser,Bovo2}. A two-dimensional F-model candidate with such defect densities could therefore be expected to show F-model behaviour except very close to the transition, where there would be a crossover to a different universality class. 

One can also envisage cases where the F-model may capture some of the physics of antiferroelectric systems  even in the absence of a direct microscopic mapping. For example, we have emphasized how the F-model illustrates a case where winding fluctuations are energetically suppressed in the ordered state and this sort of property is not necessarily confined to detailed F-model analogues, or even to two dimensional systems. Hence we consider a broad (though not exhaustive) range of experimental systems. These include spin ice thin films in a slab geometry that interpolates between two and three dimensions~\cite{Jaubert_film}, dipolar spin ice~\cite{Melko}, water ice in three dimensions and finally, artificial spin ice, a two-dimensional metamaterial~\cite{NisoliColloquium2013}. We argue that the F-model phenomenology discussed in this paper and in particular the behaviour of the energetic susceptibility is highly relevant to all of these systems and propose experimental and theoretical avenues to explore in future work. Some conclusions are drawn in Section IV.

\begin{figure*}%
    \centering
    \includegraphics[width=0.475\linewidth]{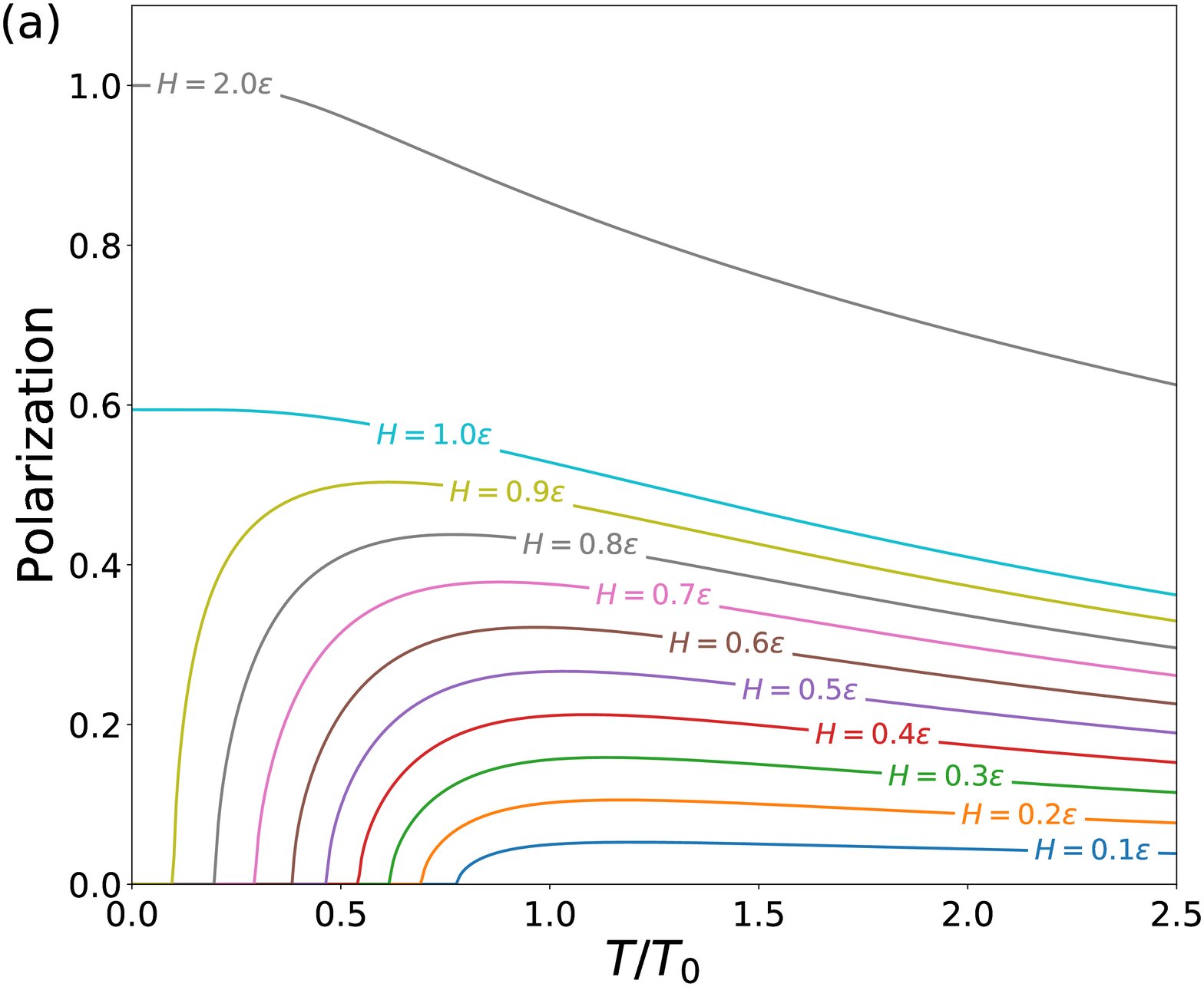} 
    \qquad
    \includegraphics[width=0.475\linewidth]{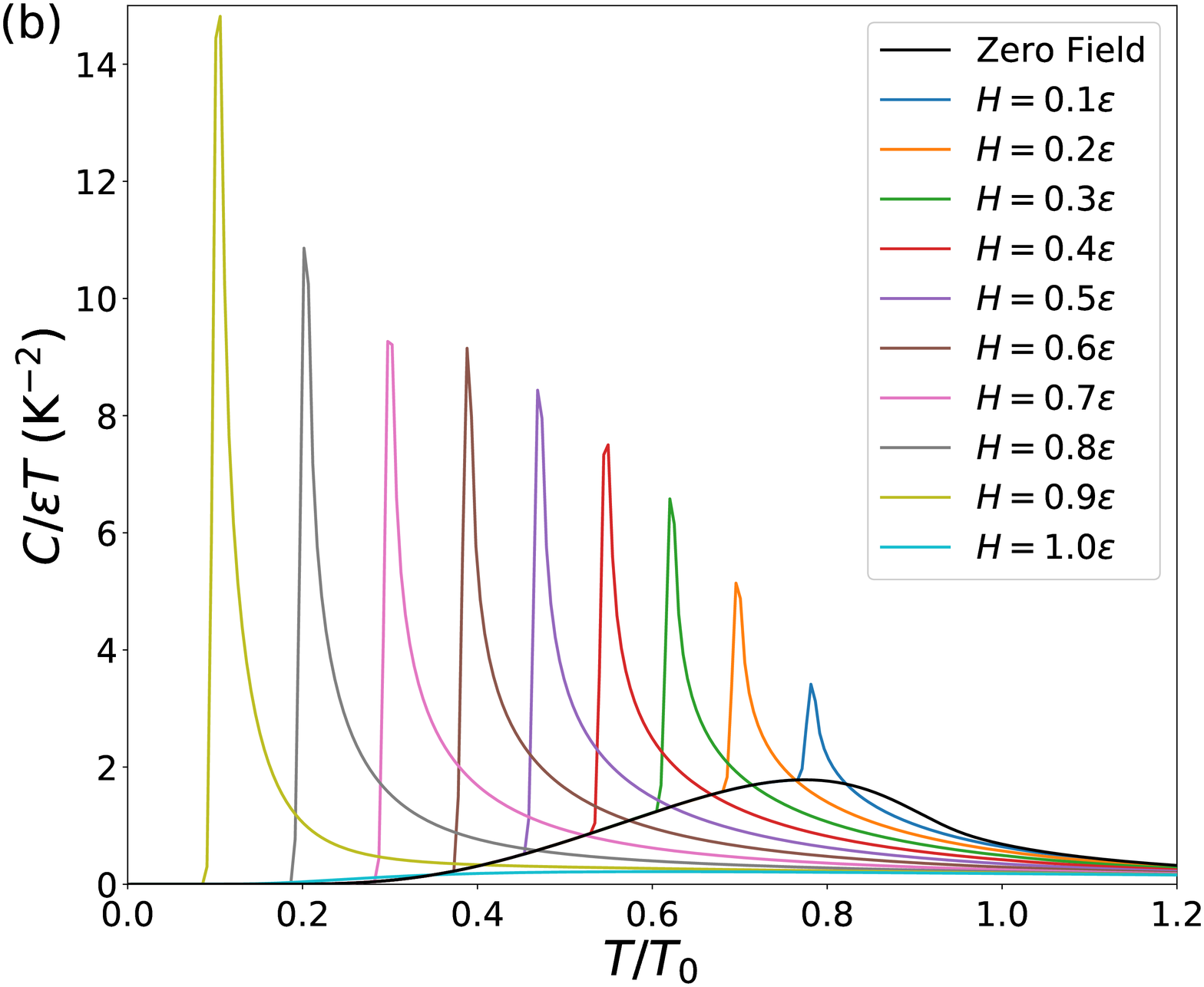} 
    \caption{(a) The polarization $m$ of the F-model as a function of the reduced temperature in varying applied fields $H$. (b) The specific heat divided by $T$ (or differential entropy increment) $C/T$ for varying fields. In applied fields $H<\epsilon$ there is a second-order phase transition at a temperature $T_{\text{C}}(H)$ as well as a maximum in the polarization at $T_{\text{max}}(H) \geq T_{\text{C}}(H)$. For fields at $H=\epsilon$ and greater (not shown), the system is aligned with the field at low temperatures and there is no phase transition.}%
    \label{fig:PolCoT-Temp}%
\end{figure*}




\section{Theory}

The following results are equally valid for the dielectric case of polarization/electric field and for the magnetic case of magnetization/magnetic field, which follows from a spin ice representation of the F-model. We generally refer to the `polarization' and `field' to cover both these cases and we give these the symbols $m$ and $H$ respectively. We further assume, without loss of generality, that $m$ and $H$ are aligned with the vertical spins of the lattice and point up with respect to the vertices in Fig. \ref{fig:vertexfiglabels}. 

\subsection{Method}

\subsubsection{Exact free energy in external field}

Neglecting external fields for the moment, the free energy of the F-model is obtained via the method of transfer matrices. Lieb's solution proceeds by recognising that the transfer matrix of the six vertex model commutes with the Hamiltonian of the Heisenberg XXZ spin chain and may be solved using the Bethe ansatz~\cite{Lieb1967,Yang1966}. Eigenvectors thus take the form of linear combinations of plane-waves with weights distributed over an even function $\rho(\alpha)$ that depends on the polarization $m \in [-1,1]$ along one of the axes via
\begin{equation}\label{Eqn: polarization}
\frac{1}{\pi}\int_{-Q}^{Q} \rho(\alpha) \text{d}\alpha = 1-m
\end{equation}
where the form of $\rho(\alpha)$ is determined by the integral equations
\begin{multline}\label{Eqn:IE_Less}
\rho(\alpha) = 
\frac{\sinh(\lambda)}{\cosh(\lambda)-\cos(\alpha)}\\
-\frac{1}{2\pi}\int_{-Q}^{Q} \frac{\sinh(2\lambda)}{\cosh(2\lambda)-\cos(\alpha-\beta)}\rho(\beta)\:\text{d}\beta\\
T<T_{0}
\end{multline}
\begin{multline}\label{Eqn:IE_Greater}
\rho(\alpha) = 
\frac{\sin(\mu)}{\cosh(\alpha)-\cos(\mu)}\\
-\frac{1}{2\pi}\int_{-Q}^{Q} \frac{\sin(2\mu)}{\cosh(\alpha-\beta)-\cosh(\mu)}\rho(\beta)\:\text{d}\beta\\
T>T_{0}
\end{multline}
with $\mu(T) = \cos ^{-1} \left(  \frac{e^{2\epsilon/T}}{2}-1 \right)$ and $\lambda(T)=\cosh^{-1}\left(\frac{e^{2\epsilon/T}}{2}-1 \right)$ and the bounds on the integrals $\pm Q$ determined by minimising the free energy per vertex
\begin{equation}\label{Eqn:freeE_ZF}
f(Q,T)=-\frac{T}{2\pi} \int_{-Q}^{Q} \ln\left(L(\alpha) \right) \rho(\alpha) \text{d}\alpha
\end{equation}
where
\begin{equation}\label{Eqn:L}
L(\alpha) = 
\begin{cases}
(e^{\lambda}-e^{-(\lambda+i\alpha)})/(e^{-i\alpha}-1)&
T<T_{0}\\
(e^{i\mu}-e^{\alpha-i\mu})/(e^{\alpha}-1)&
T>T_{0}
\end{cases}
\end{equation}

While closed-form solutions exist for the free energy in zero field (see \cite{Baxter}), when an external field $H$ is applied along one of the F-model axes solutions must be found numerically by determining the distribution function $\rho(\alpha)$ and corresponding $Q$ that minimise
\begin{equation}\label{Eqn:freeE_H}
f(Q,T,H) = f(Q,T) - m(Q)H
\end{equation}

In this work the integral equations \ref{Eqn:IE_Less} and \ref{Eqn:IE_Greater} were solved numerically for candidate values of $Q$ using the Nystr{\"o}m method \cite{Nystrom1930,Delves1985,Press2007}, in which the Gauss-Legendre quadrature rule is applied to the integrals to obtain a system of $n$ simultaneous equations for $\rho(\alpha)$ evaluated at the $n$ quadrature points $\alpha_{i} \in [-1,1]$, giving (e.g. for equation \ref{Eqn:IE_Less})
\begin{multline}\label{Eqn:IE_Less_GaussLegendre}
\rho(Q \alpha_{i}) =
\frac{\sinh(\lambda)}{\cosh(\lambda)-\cos(Q \alpha_{i})}\\
-\frac{Q}{2\pi}\sum_{j=0}^{n-1} \frac{w_{j} \sinh(2\lambda)}{\cosh(2\lambda)-\cos(Q \alpha_{i}-Q \beta_{j})}\rho(Q \beta_{j})
\end{multline}
where $w_{j}$ are the quadrature weights and $\beta_{j}$ are the same set of quadrature points as $\alpha_{i}$. Once the values of $\rho(Q\beta_{j})$ are known, we may replace $\alpha_{i}$ in equation \ref{Eqn:IE_Less_GaussLegendre} with arbitrary $\alpha$ to interpolate $\rho(\alpha)$ between the quadrature points.

By finding the $Q$ that minimises the free energy (equation \ref{Eqn:freeE_H}) for given $T$ and $H$, it is thus possible to determine the equilibrium values of experimentally relevant quantities such as the the polarization, susceptibility and specific heat. The results presented below were calculated using $n=100$ quadrature points.

In its antiferromagnetic phase, the free energy of the F-model is independent of any applied direct fields since there is no net polarization $m$ for the fields to couple to (see Eq. \ref{Eqn:freeE_H}). External fields therefore do not affect any thermodynamic quantities provided they are not strong enough to take the system out of the antiferromagnetic phase. The spontaneous staggered polarization $m^{\dagger}$ (the order parameter of the antiferromagnetic phase) obtained via the Bethe ansatz
\begin{equation}\label{Eqn:StagPol}
    m^{\dagger} = \prod^{\infty}_{n=1} \tanh^{2}(n \lambda),
\end{equation}
with $\lambda$ defined as for Eq. \ref{Eqn:IE_Less}, is then valid for all small fields (the precise field at which the system leaves the antiferromagnetic phase is addressed in the Results section).

Unfortunately the Bethe ansatz is only valid under conditions that require the two antiferromagnetic ground states $m^{\dagger}=\pm 1$ to be degenerate \cite{Baxter1971}. It is therefore not possible to obtain results for the response of the F-model to staggered fields that break this degeneracy, though Baxter \cite{Baxter1973} suggests, based on scaling theory arguments, that as the transition temperature is approached the staggered susceptibility strongly diverges as
\begin{equation}\label{Eqn:StagSuscScale}
    \chi^{\dagger} \sim \lambda^{-2} \exp (\pi^{2}/2\lambda)
\end{equation}
where again $\lambda$ is defined as for Eq. \ref{Eqn:IE_Less}. This is as much as Bethe ansatz methods can tell us about the staggered susceptibility, though further insight can be gained by studying the F-model on a Bethe lattice as discussed in the following section.

\subsubsection{Cavity method}

Bethe lattice methods have been widely used to obtain exact results in statistical physics including in the study of the Ising model~\cite{Baxter}, Potts models~\cite{Peruggi1983} and spin glasses~\cite{Mezard2001}. The methods may be regarded as generalisations of mean field methods in which near-neighbour correlations are taken into account by placing a system on an infinite, acyclic connected graph on which all vertices have the same valence and studying the response of a central element (which may be a spin, a vertex or a larger unit) in the ``bulk'' of the graph to the system surrounding it. It is assumed that while each site is correlated with its nearest-neighbours there is no correlation between any of the neighbours so that the graph may be separated into uncorrelated sub-graphs by removing one of its edges. Such methods have been successfully applied to vertex models \cite{Foini2013} and to spin ice systems~\cite{Yoshida2002,Jaubert} for which they provide excellent agreement with the experimentally observed temperature-dependence of the magnetic susceptibility.

In this work we develop the cavity method approach of Foini et al. \cite{Foini2013} to obtain approximate expressions for the staggered susceptibility of the F-model. This method works by expressing the partition function in terms of self-consistent probabilities for the four spins of the central vertex having certain orientations and identifying fixed points that correspond to thermodynamic phases. Remarkably, this yields the exact same phase diagram as Lieb's solution of the square-lattice six-vertex model.

The (staggered) susceptibility over a particular phase may be determined by defining a stability matrix $\mathsf{M}$ as the Jacobian matrix of the self-consistent probabilities evaluated at a particular fixed point, where
\begin{equation}\label{Eqn:BetheSuscVert}
    \chi^{\dagger} \simeq \sum_{r=1}^{\infty} \Tr \mathsf{M}^{r}
\end{equation}
Clearly, unless all the eigenvalues of $\mathsf{M}$ have absolute values less than unity, Eq. \ref{Eqn:BetheSuscVert} diverges and the fixed point at which $\mathsf{M}$ was evaluated is unstable. The stability matrix thus also provides a criterion for determining under what conditions the different thermodynamic phases are stable.

We adapt the cavity method of Ref. \cite{Foini2013} by calculating the appropriate stability matrix for a Bethe lattice of F-model vertices (referred to as a ``vertex tree'' from now on). Hence we obtain the staggered susceptibility in the antiferromagnetic phase as 
\begin{equation}
    \chi^{\dagger}_{\rm{Vert}}(\theta) \simeq \frac{2}{4^{\theta-1}-1}
\end{equation}
where $\theta^{-1} = T/T_0$.

This approximation may be improved by carrying out the same calculation for a Bethe lattice of plaquettes containing four F-model vertices (a ``plaquette tree''), instead of for a vertex tree. In this case, we obtain the rather more complicated expression
\begin{multline}\label{Eqn:BetheSuscPlaq}
    \chi^{\dagger}_{\rm{Plaq}}(\theta) \simeq 
    \frac{4-3(2^{2 \theta +1})}{3(2^{2 \theta +1})-2^{4 \theta +1}+64^{\theta }-4} \\ 
    -\frac{4^{2-3 \theta }}{3} +\frac{29( 4^{1-2 \theta })}{9}-\frac{253\times 4^{-\theta }}{27} \\
    +\frac{7}{4^{\theta }-4}+\frac{64}{27 \left(4^{\theta }-3\right)}
    +\frac{2}{4^{\theta }-2}\\
    +\frac{1}{2+16^{\theta }-2^{2 \theta +1}}-\frac{2^{4 \theta +1}+3\times 4^{\theta }-2}{3(2^{2 \theta +1})+64^{\theta }-4}
\end{multline}

Equations \ref{Eqn:BetheSuscVert} and \ref{Eqn:BetheSuscPlaq} provide approximate results for the temperature dependence of the staggered susceptibility in the F-model. It should be noted, however, that since the cavity method neglects long-range correlations it cannot capture the topological features of the F-model and the transition becomes second-order. The effect of successively longer-range correlations could be determined by studying Bethe lattices with larger plaquettes, but we do not explore these higher-order corrections in this work.


\subsection{Results}

\subsubsection{Polarization and specific heat as function of $T$}

We begin by discussing the polarization and specific heat of the F-model with a field $H$ applied along one of its axes. The immediate effect of applying an external field is to convert the infinite order phase transition at $T_{\rm 0}$ into a second order antiferromagnetic transition at $T_{\rm C}(H)<T_{\rm 0}$ below which the polarization vanishes. Figure \ref{fig:PolCoT-Temp}a shows the exact polarization as a function of temperature in fixed external fields up to $H \gtrsim \epsilon$. The direct polarization is finite only in applied field and zero for $H < \epsilon$ and $T < T_{\rm C}(H)$. Hence, as anticipated, for all $H < \epsilon$ the low temperature phase is a vacuum for system-spanning polarization windings owing to the fact that they cost energy of order the system size. 

Although the transition in applied field is second order, there is one aspect that marks it out as a very unusual second order transition: remarkably, the free energy at temperatures below $T_{\rm C}(H)$ is identical to that in zero field. This arises because when $m=0$, Eq. \ref{Eqn:freeE_H} reduces to Eq. \ref{Eqn:freeE_ZF}, so that while there are significant differences on the high-temperature side of the transition the specific heat collapses onto the zero-field curve for $T<T_{\rm C}(H)$ (Fig. \ref{fig:PolCoT-Temp}b). This again is a consequence of all polarization being carried by system-spanning windings of divergent energy cost, which is an unusual property of the F-model.  

\subsubsection{Polarization as function of $H$}

\begin{figure}
	\centering
	\includegraphics[width=1.0\linewidth]{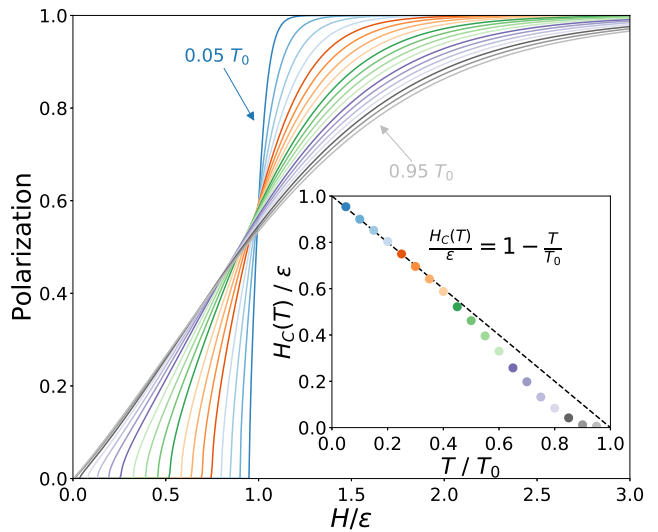}
	\caption{The F-model polarization as a function of applied field at various temperatures between $0.05T_{0}$ and $0.95T_{0}$. For small temperatures the critical field $H_{\text{C}}$ is proportional to $|T-T_{0}|$ (inset).}
	\label{fig:Pol-H}
\end{figure}

The polarization is also shown for fixed temperatures as the external field is swept (Fig. \ref{fig:Pol-H}). Reminiscent of the temperature sweeps, the polarization stays fixed at zero until a critical field $H_{\rm C}(T)$ is reached and then it  abruptly (though continuously) becomes finite-valued. In the limit $T \to 0$ the transition becomes step-like with $m=0$ for $H<\epsilon$ and $m=1$ for $H>\epsilon$, with the critical field decreasing as $H_{\rm C}(T) = \epsilon - T \ln 2$ for small finite $T$ (inset, Fig. \ref{fig:Pol-H}), a feature reproduced in a Faraday loop representation of the F-model by Nisoli \cite{Nisoli_preprint} discussed further below.
The step-like increase in polarization is again easily anticipated by considering the in-field energy of a system-spanning windings: as $T\rightarrow 0$, they cost infinite energy below $H_{\rm C}$ and return infinite energy above, hence the step response.    

It is interesting, however, that if the field is fixed exactly to $H = \epsilon$ (Fig. \ref{fig:PolCoT-Temp}a), then the polarization reaches a zero temperature limit of neither zero nor unity, but instead $m \approx 0.6$. This occurs because when $H = \epsilon$, the ground state vertices (types 5 and 6 in Fig. \ref{fig:vertexfiglabels}) and type 1 and 4 vertices, whose vertical spins are aligned with the field, become degenerate in energy so that when $T \to 0$ the system is in a disordered phase with a mixture of these four vertices. Surprisingly, the relative amounts of ground state vertices and vertices of types 1 and 4 (and hence the value of $m$) seems to depend on what boundary conditions are used even in the thermodynamic limit, with $m \approx 0.6$ only appearing as the zero-temperature limit when periodic boundary conditions are employed. In general the polarization as a function of $H$ at zero temperature is a step function with $m=0$ for $H<\epsilon$, $m=1$ for $H<\epsilon$ and with $m$ taking some intermediate value for $H=\epsilon$.

\subsubsection{Polarization maximum}

Another clear feature of Figure \ref{fig:PolCoT-Temp}a is a maximum in the polarization at some temperature $T_{\rm max}(H) \geq T_{\rm C}(H)$. This maximum arises as follows. Since in finite fields, the polarization increases from $m=0$ when the temperature rises above $T_{\rm C}(H)$ but must tend to zero as $T \to \infty$, where system-spanning windings freely fluctuate and a Curie law is obeyed, the polarization must be non-monotonic in temperature. The consequent maximum in the polarization at a temperature higher than the critical temperature is not so unusual, being a rather general characteristic of low-dimensional antiferromagnets that stems from antiferromagnetic correlations that build up as the transition temperature is approached from above, thereby reducing the statistical weight of direct polarization fluctuations \cite{Sykes1962, Domb1996}.

\subsubsection{Evolution of $T_{\rm C}$ and $T_{\rm max}$ with $H$}

\begin{figure}
	\centering
	\includegraphics[width=1.0\linewidth]{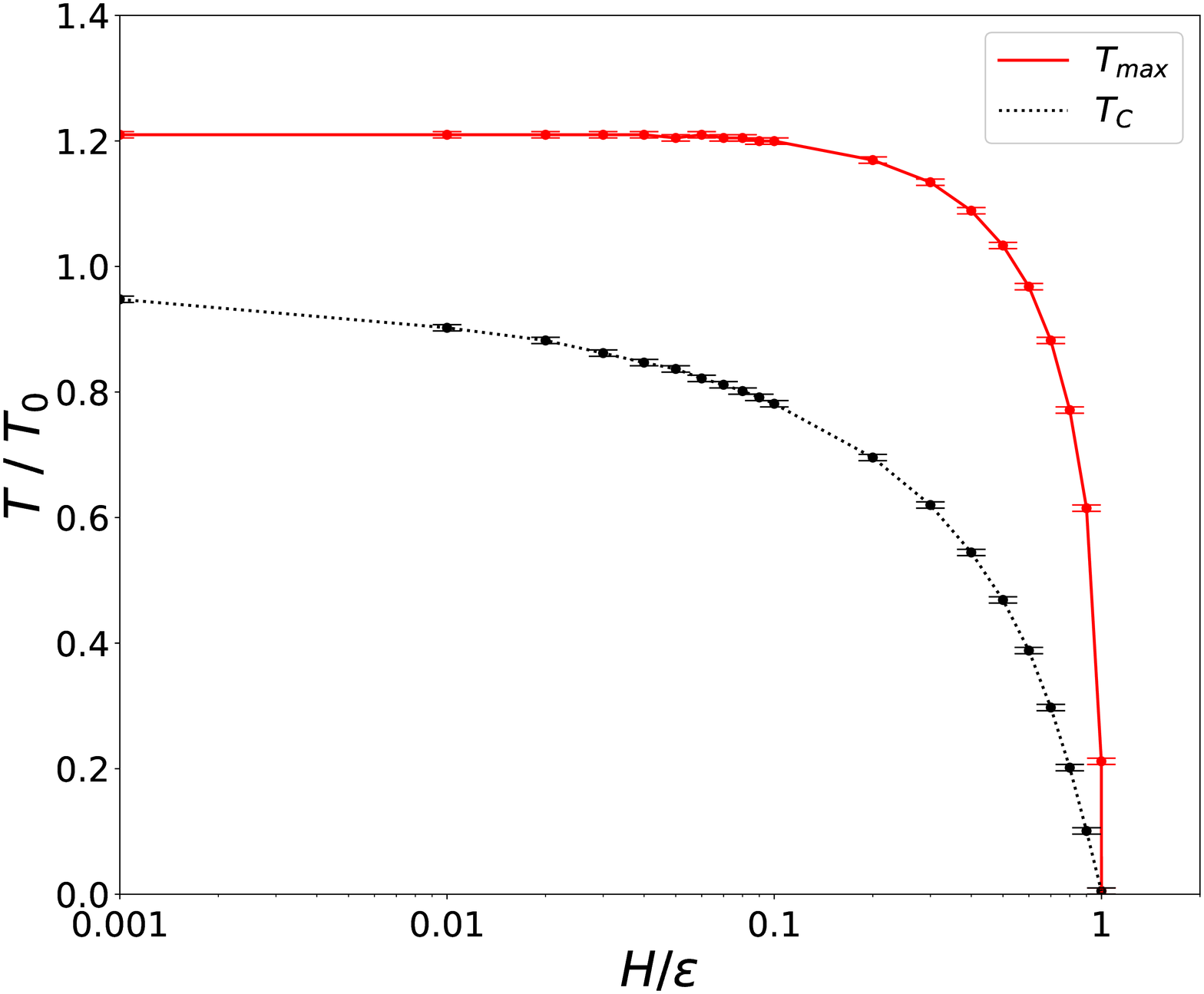}
	\caption{The temperatures $T_{\text{C}}$ and $T_{\text{max}}$ as functions of the applied field. In the limit of very small fields $T_{\text{C}} \to T_{0}$ and $T_{\text{max}} \to \; \sim 1.213 \, T_{0}$, while in the limit $H \to \epsilon$ the two temperatures converge to zero.}
	\label{fig:TMaxC}
\end{figure}

It is interesting to follow the evolution of the critical temperature $T_{\rm C}(H)$ and the temperature at which the polarization reaches its maximum value $T_{\rm max}(H)$ as the field is decreased from $H=\epsilon$ to zero, shown in Figure \ref{fig:TMaxC}. At $H=\epsilon$ (we will take the field to lie along the upwards vertical direction of Fig. \ref{fig:vertexfiglabels} as before), vertices 1, 4, 5 and 6 become degenerate and the antiferromagnetic ordered phase vanishes so that $T_{\rm C} = T_{\rm max} = 0$. Reducing the field increases both $T_{\rm C} $ and $T_{\rm max}$ as the antiferromagnetic phase reasserts itself, with $T_{\rm C} \to T_{0}$ as the field vanishes. The temperature of the polarization maximum $T_{\rm max}$ is greater than $T_{\rm C}$ for all $H<\epsilon$ with a limiting value of $T_{\rm max} \sim 1.213\;T_{0}$ as $H \to 0$. 

\subsubsection{Energetic susceptibilty}

The energetic susceptibility in zero field may be defined as $\chi T/C$(where $C$ is the Curie constant), or equivalently as $\chi(T)/\chi(\infty)$, presuming a Curie law holds in the high temperature limit. Here a Curie law does arise from the free fluctuations of the system-spanning polarization loops at sufficiently high temperature. As  shown by Macdonald et al.~\cite{Macdonald} in their study of the kagome ice phase of spin ice, the energetic susceptibility is a sensitive measure of topological sector fluctuations, while for an ordinary antiferromagnet, Fisher~\cite{Fisher1962, deJonghMiedema} showed it is closely related to the internal energy $U$. For greater generality, we define the energetic susceptibility here as 
\begin{equation}
\tilde{\chi} = \chi/\chi_{\rm c}    
\end{equation}
where $\chi_{\rm c}$ is the susceptibility of the freely fluctuating six vertex state, here reached in the high temperature limit. 

\begin{figure}
	\centering
	\includegraphics[width=1.0\linewidth]{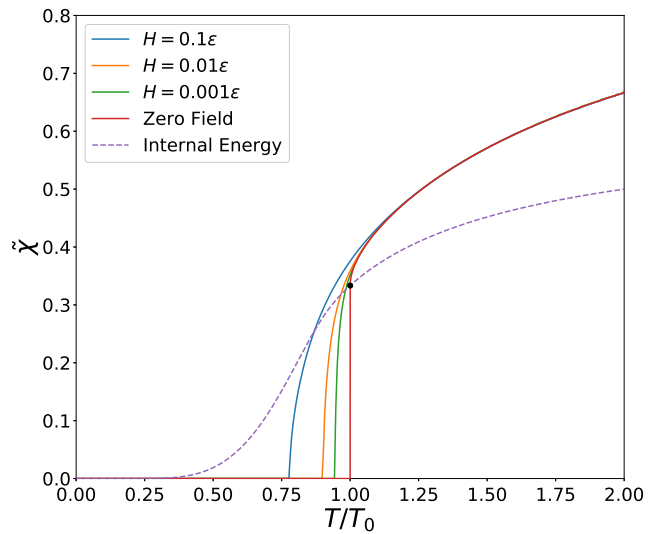}
	\caption{The energetic susceptibility as a function of temperature for different applied fields (with the susceptibility calculated as $\chi = m/H$) and in zero field (using the exact result of equation \ref{Eqn:ExactChiToC}) together with the exact zero-field internal energy. At the transition temperature $T_{0}$, the internal energy and the zero-field energetic susceptibility intersect at $U = \tilde{\chi} = 1/3$ (indicated by a black circle).}
	\label{fig:ChiToC}
\end{figure}

Figure \ref{fig:ChiToC} displays the energetic susceptibility as a function of temperature in different fields, taking $\chi \approx m/H$ in the small-field limit, together with the exact result \cite{LiebWu1972} 
\begin{equation}\label{Eqn:ExactChiToC}
   \tilde{\chi} = \frac{\pi}{3(\pi - \mu(T))}
\end{equation}
with $\mu(T)$ defined as it was for Eq. \ref{Eqn:IE_Greater} above.  Also shown in Figure \ref{fig:ChiToC} is the exact energy $U(T)$, shifted to make $U(0) = 0$. We see that the energetic susceptibility evolves to an abrupt jump at the transition as topological sector fluctuations are suppressed. This is reminiscent of the results of Macdonald et al. for the case of `kagome ice'~\cite{Macdonald}, although in that case, the suppression is dynamical, rather than energetic. There is however, a great difference with the internal energy, which remains finite below $T_0$ on account of the excitation of closed, non polarized loops. This complete  breakdown of the Fisher relation arises precisely because in the F-model there are no local, polarized excitations, and we can clearly promote breakdown of the Fisher relation as a signature of the dominance of topological sector fluctuations. Rather intriguingly, with the shifted $U$, the relation $U(T_{0})=\tilde{\chi} = 1/3$ is exact at the critical temperature. 

\subsubsection{Faraday loop representation}

It is interesting to compare our results to those of a very recent work by Nisoli \cite{Nisoli_preprint} that makes the topological aspects of the F-model explicit in a ``Faraday loop'' representation and seeks to derive the behaviour of the system directly from its topology. In the loop representation employed, the phase space of the F-model may be partitioned into the topological sectors $\mathcal{T}$ (all configurations with no system-spanning loops), $\mathcal{W}_{0}$ (all configurations with system-spanning loops but no net polarization), and $\mathcal{M}$ (all configurations with a net polarization). Nisoli postulates by analogy with the BKT transition that, at low temperatures, the system is asymptotically confined to the sector $\mathcal{T}$ while at high temperatures the system is asymptotically confined to $\mathcal{W}_{0}$, with the zero-field transition consisting of a transition between the two topological sectors. Impressively, this picture allows Nisoli to reproduce qualitatively the behaviour of the magnetic and antiferromagnetic order parameters as a function of field and temperature, as well as determine exactly the transition temperature and behaviour of $H_{\rm C}(T)$ in the limit $T \to 0$.

It is nevertheless important to make a distinction between the asymptotic confinement of the low-temperature phase to $\mathcal{T}$, which rigorously excludes topological sector fluctuations and therefore has $\chi=0$, and the asymptotic confinement of the high-temperature phase to $\mathcal{W}_{0}$, which has a finite susceptibility for all $T$. In the latter case, while fluctuations between topological sectors vanish at the thermodynamic limit $N = \infty$ they remain significant for large but finite systems.

It is clear moreover that, in contrast to the case of asymptotic confinement at the thermodynamic limit, for a large but finite system the analogy with the BKT transition is not valid in detail. As remarked in the Introduction, and shown in Ref. \cite{Faulkner}, the BKT transition is a dynamical ergodicity breaking, where charge confinement suppresses topological sector fluctuations (equivalent to the polarized loops of Ref. \cite{Nisoli_preprint}). There is no equivalent suppression in the F-model because there are no charge dynamics, and if charges were introduced, they would not be confined. Therefore, $\mathcal{W}_0$ does not represent the fixed-field ensemble at $T > T_0$, which necessarily involves the polarized states $\mathcal{M}$. For this reason we prefer to speak of confinement to $\mathcal{W}_{0}\cup\mathcal{M}$, which is topologically protected in the same way as the low-temperature phase.

\subsubsection{Universal jump}

The jump of the the energetic susceptibility to zero at $T_0$ is reminiscent of the universal jump in spin stiffness at the BKT transition~\cite{NK}, and indeed, there is a certain connection. The universal jump in the BKT system (say the 2D-XY model) is equivalent to the magnetization correlation exponent $\eta$ jumping from 0.25 to zero, with increasing temperature. Here we note that the  correlation exponent $\eta(T)$ of the staggered order parameter of the F-model is believed (see Ref. \cite{Weigel}) to obey a relation that makes it an exact multiple of our $\tilde{\chi}$:
\begin{equation}
\tilde{\chi} =\eta/3.
\end{equation}
According to the exact equations, the jump then occurs at $\eta(T_0) = 1$ or $\tilde{\chi} =1/3$ (which is different numerically to the XY model case~\cite{Weigel}).  Below the ordering temperature, $\eta$ falls abruptly to zero, order parameter correlations become exponential and the staggered susceptibility $\chi^{\dagger}$ becomes finite.

\subsubsection{Coulomb phase correlations}

As the temperature increases above $T_0$, the energetic susceptibility gradually reaches its high temperature limit of unity.  The exponent $\eta$ thus becomes equal to $3$ in the high temperature limit. This is different to the two dimensional dipolar exponent ($= 2$) of the polarization correlations~\cite{YAM1980}. The difference is a significant one as with $\eta = 3$, a two-dimensional integral over correlations converges,  while with $\eta = 2$ it converges only conditionally. Hence the effective interaction is much longer range in the case of the direct polarization, emphasising the essentially ferroelectric (or ferromagnetic) nature of the correlations in the six vertex state. 

As the temperature is lowered, the roles of ferroelectric and antiferroelectric correlations are reversed as the exponent evolves from $\eta= 3$  in the high temperature limit, to $\eta(T_0) = 1$ at the transition. The order parameter susceptibility $\chi^\dagger$ obeys the equation:
\begin{equation}\label{Eqn:ChiT-Eta}
\chi^\dagger T  \sim  \int r^{1-\eta} \:\text{d}r \sim L^{2-\eta},   \end{equation}
(where $r$ is distance and $L$ is system size). 
The order parameter susceptibility therefore diverges in the range $\eta = 1$ to $\eta= 2$, but above $T_2  = 2 T_0$ it is non-divergent.  This special temperature has been discussed in reference to the BCSOS model representation of the F-model \cite{Mazzeo1992, Mazzeo1994}. 

From this we see that the energetic susceptibility $\tilde{\chi}$ measures not only topological sector fluctuations, but also the correlations of the hidden antiferroelectric order parameter, the growth of which eventually destroys the Coulomb phase. Indeed, it seems that a criterion for stability of the Coulomb phase is $\tilde{\chi} \ge 1/3$ and this is potentially more general than for the F-model, as discussed subsequently.

\subsubsection{Staggered polarization and susceptibility in the antiferromagnetic phase}

\begin{figure}
	\centering
	\includegraphics[width=1.0\linewidth]{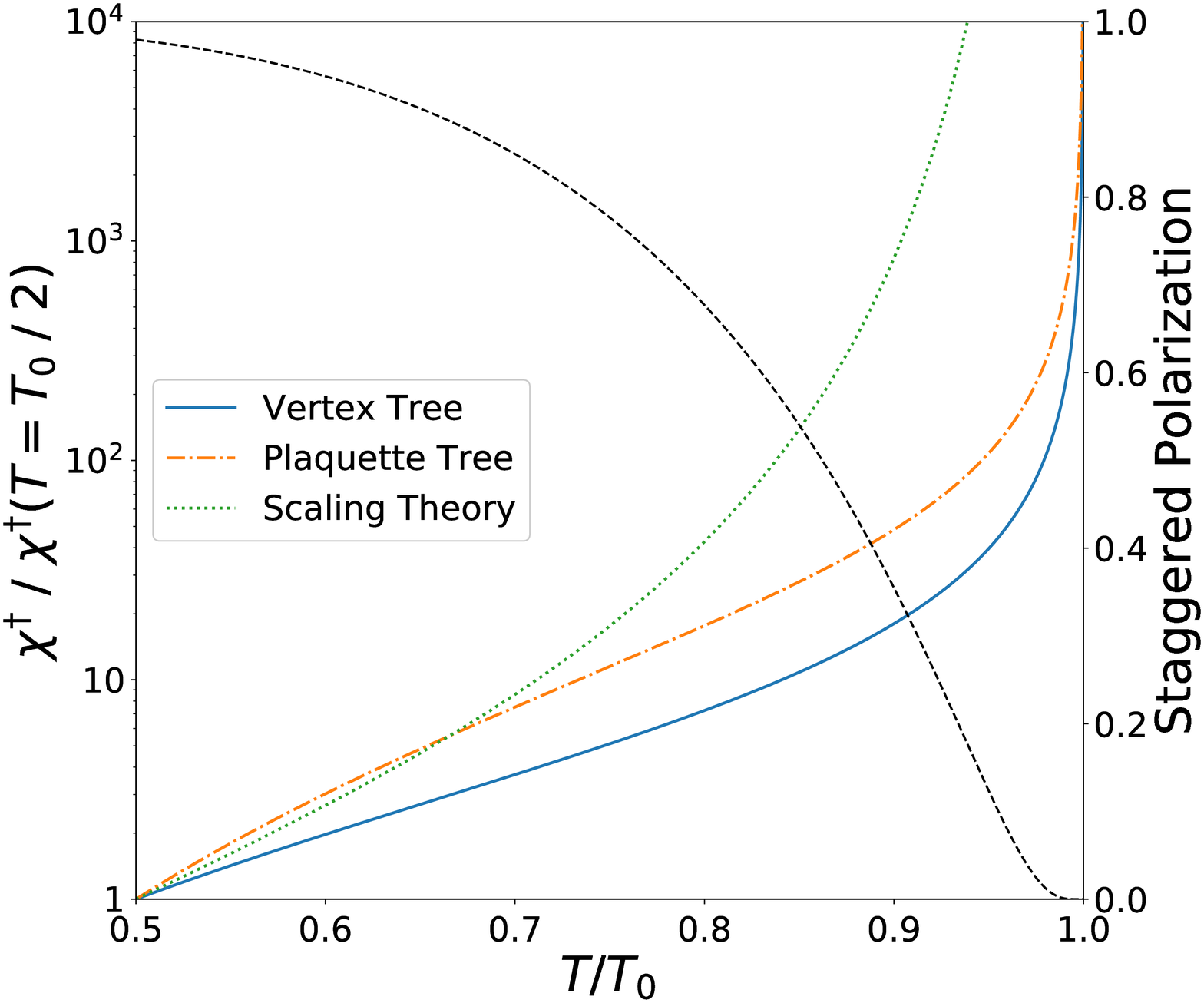}
	\caption{The staggered susceptibility $\chi^{\dagger}$ of the F-model on a vertex and plaquette tree, together with the asymptotic staggered susceptibility predicted for temperatures close to $T_{0}$ by scaling theory and the exact result for the staggered polarization. In all three cases, the staggered susceptibility diverges as the transition temperature is approached.}
	\label{fig:StaggeredSusc}
\end{figure}

We finally comment on the temperature dependence of the staggered polarization and susceptibility of the F-model in its antiferromagnetic phase, shown in Figure \ref{fig:StaggeredSusc}. The staggered polarization (Eq. \ref{Eqn:StagPol}) is equal to unity at $T=0$ and decays slowly until the transition temperature where it vanishes infinitely continuously through the infinite-order F-model transition. Both the vertex tree (Eq. \ref{Eqn:BetheSuscVert}) and the plaquette tree (Eq. \ref{Eqn:BetheSuscPlaq}) give a staggered susceptibility that tends to zero as $T \to 0^{+}$ and diverges as a power law as $T \to T_0$, in contrast to the essential singularity (Eq. \ref{Eqn:StagSuscScale}) predicted for the square-lattice F-model from scaling theory.

As noted in the section on the cavity method above and by Foini et al. \cite{Foini2013} it is remarkable that though the order of the transition is changed, both the vertex tree and plaquette tree have the same transition temperature as the square-lattice F-model. In fact, the F-model also has the same transition temperature on a kagome lattice \cite{Baxter} and more generally it has been shown that, provided certain constraints are met, all regular planar graphs with valence 4 have the same F-model transition temperature~\cite{Baxter1978}. It would be interesting to determine whether the F-model on the pyrochlore lattice, which has valence 4 but is non-planar, shows the same critical plane.

\section{Application to experimental systems}

\subsubsection{Spin Ice Thin Films}

Spin ice thin films of ${\rm R_2Ti_2O_7}$ (R = Dy, Ho) have been developed both on commercial substrates~\cite{Leusink, Barry} and on the isomorphous pyrochlore substrate ${\rm Y_2Ti_2O_7}$~\cite{Bovo1, Bovo2}. In the latter case, which we shall consider, the epitaxial strain is homogeneous even in relatively thick films and the six vertex degeneracy of spin ice is strongly broken, mainly by strain induced changes in exchange interactions~\cite{Bovo1}. The degeneracy breaking apparently mimics that of the F-model and a component of the specific heat in these experimental systems closely approximates the exact zero-field result with $\epsilon \approx 0.55$  K~\cite{Bovo2}. An F-model type transition is further observed in numerical simulations based on the dipolar spin ice model in slab geometry~\cite{Jaubert_film}.

\begin{figure}
	\centering
	\includegraphics[width=1.0\linewidth]{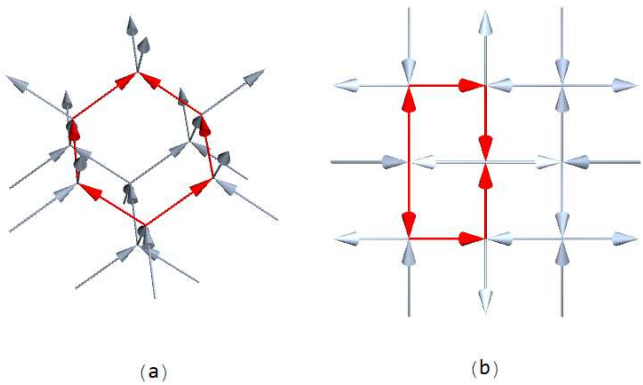}
	\caption{(a) The F-model ground state on a section of the pyrochlore lattice. In pyrochlore systems the staggered polarization in the (100) plane is proportional to the direct polarization along the [100] direction. (b) The same state projected onto the (100) plane. Fields in the (100) plane couple to a vertical and horizontal sublattice as for the square-lattice F-model. Note that on the pyrochlore lattice the minimal closed loop contains six spins (an example is highlighted in red).}
	\label{fig:PyrochloreProjection}
\end{figure}

In detail, there is a geometrical difference between spin ice slabs and the F-model, whereby the spins (arrows) of spin ice are tilted out of plane (Fig. \ref{fig:PyrochloreProjection}). The tilting itself is of minor importance as the in-plane magnetization remains analogous to the F-model polarization while the out-of-plane magnetization becomes analogous to the staggered order parameter. However, an idealized monolayer slab of epitaxially strained spin ice (Fig. \ref{fig:PyrochloreProjection}) does not map exactly to the F-model, because there is always some surface termination of connections. The simulations~\cite{Jaubert_film} consider few-monolayer systems with the dipolar spin ice Hamiltonian containing both dipolar and exchange terms. They focus on hypothetical unstrained and weakly strained systems, in which, as the temperature is lowered, there is first an ordering of surface spins or arrows, and  then an F-model type transition at lower temperature. It is not clear how applicable these results are to the real spin ice films, which are strongly strained and should be much closer to the F-model than the systems considered in Ref. \cite{Jaubert_film}. 

A spin ice slab of a few monolayers does, however, have the same local structure as the F-model, and a similar kind of in-plane connectivity. In its simplest description, the near neighbour spin ice model, the F-transition is strongly obscured by the excitation of ice rule breaking defects which  introduce an exponential cut-off into the integrals of Equation \ref{Eqn:ChiT-Eta} and prevent the divergence of the order parameter susceptibility between $T_0$ and $T_2$.  However, it seems that in the real systems~\cite{Callum}, additional dipole interactions push ice rule breaking vertices to an energy much greater than $\epsilon$, making them largely negligible on the scale of $T \sim \epsilon$ and restoring the F-model behaviour observed in experiment. Hence it would be very  interesting to further compare the experimental polarization and finite field specific heat with the F-model results derived here.  

There then arises the question of how these properties will evolve as the film thickness is increased, to eventually reach the limit of three dimensions. The related problem of bulk spin ice homogeneously compressed out of plane has been studied in Ref. \cite{Jaubert_pressure} where an interesting phase transition between a Coulomb liquid and a fully magnetized phase was discovered. This scenario is related to the Slater KDP model~\cite{Slater}, where the four magnetized vertices (1-4 in Fig. \ref{fig:vertexfiglabels}) lie lowest in energy, rather than to the F-model, where the two unmagnetized vertices (5 and 6) form the ground state. In the spin ice geometry, for a single layer, it would correspond to in-plane ferromagnetism.  The authors of Ref. \cite{Jaubert_pressure} did not consider the F-model case, stating that it could not be experimentally achieved. In fact, despite impressive developments in physical pressure experiments on spin ice~\cite{Edberg2019,Henelius}, the availability of epitaxial strain renders the F-model case of homogeneous in-plane compression the more accessible of the two scenarios. Therefore it would be very interesting to revisit the work of Ref. \cite{Jaubert_pressure} to consider the F-model case. 

\subsubsection{Dipolar spin ice}

Many three-dimensional ice rule systems, like water ice and spin ice, enter an approximate six vertex state or Coulomb phase at low temperature, before they eventually order (if dynamically possible) into an antiferroelectric or antiferromegnetic state at a lower temperature, $T_0$.  Since ice-rule defects are effectively excluded at such low temperatures, polarization can once again only occur by the excitation of system-spanning windings, and these systems can potentially be represented to some extent by the F-model, despite their higher dimensionality.

Dipolar spin ice is a model of bulk spin ice that orders antiferromagnetically off an effective `Pauling plateau' with six vertex correlations~\cite{Melko}. It is considered to be an accurate and realistic model of real spin ices even though the ordering transition is not observed in those systems, probably because of dynamical arrest. Ryzhkin has also formulated a pseudo-spin model of water ice that is equivalent to dipolar spin ice \cite{Ryzhkin84}.

In Ref. \cite{Bovo_special} there are presented several simulations of generalised dipolar spin ice, with different coupling parameters, where it was noted that the energetic susceptibility, defined there as $\chi T/C$ (where $C$ is the Curie constant), invariably jumps discontinuously from about 2/3 to zero at the ordering transition. This quantity $\chi T/C$ contains spin orientational factors which affect its numerical value: in dipolar spin ice our $\chi_{\rm c}$ for the six vertex model takes the (non-universal) value $\chi_{\rm c} = 2$. Division of $\chi$ by $\chi_{\rm c}$ removes the orientational factors and yields our energetic susceptibility $\tilde{\chi}$. In dipolar spin ice, the energetic susceptibility $\tilde{\chi}$, as we have defined it, then jumps from 1/3 to zero at $T_0$ -- precisely the same jump that occurs in the  two-dimensional F-model.

The ordering transition in dipolar spin ice is first order, so not at all the same as the infinite order transition of the F-model. Nevertheless, in view of this result, it seems reasonable to propose  that the criterion $\tilde{\chi} = 1/3$ marks the limit of stability of the Coulomb phase, regardless of dimension. As found in Ref. \cite{Bovo_special}, the growth of antiferromagnetic correlations is signalled by a slow decline in the energetic susceptibility below its six vertex value.

\subsubsection{Water ice}

Since an apparent universal jump $\tilde{\chi} \approx 1/3 \rightarrow \tilde{\chi} = 0$ is observable in dipolar spin ice \cite{Bovo_special}, and an equivalent model has been shown by Ryzhkin \cite{Ryzhkin84} to be an appropriate model of water ice, it is most tempting to check for the universal jump in experimental data at the ordering transition from lightly doped ice Ih to ordered ice XI at $70$ K. We refer to the very early data of Kawada \cite{Kawada1972}, who observed a jump to zero in dielectric constant $\epsilon' \approx 1 + \chi$ at the ordering transition of 70 K. If we assign the peak in $\epsilon'$ at 78 K for the lowest measuring frequency (1 Hz) as the dielectric constant of the 6-vertex phase, then the observed jump in $\tilde\chi$ is indeed consistent with the value of 1/3. However the jump does evolve with applied frequency, so this result would need to be checked for full confidence. 

Another potential realisation of the F-model is ice-II~\cite{Shepard}, a proton-ordered antiferroelectric phase of ice. While it does not necessarily map microscopically on to the F-model, it does exhibit one distinctive F-model property, illustrated in Fig. \ref{fig:Pol-H}: the completely `frozen' nature of the system below the transition, which arises because polarized states have energy at least of order $L$. In ice-II this is indicated by a remarkable absence of polarization relaxation~\cite{Whalley}. However, ice-II decomposes into other ice phases before any transition occurs. 

\subsubsection{Artificial Spin Ice}

Artificial spin ice (ASI) systems are geometrically frustrated magnetic metamaterials built up out of single-domain ferromagnetic nanoislands arranged in two-dimensional arrays. When the first ASI system was fabricated on a square lattice \cite{Wang2006} it was quickly noted that due to inequivalent interactions between the opposite and perpendicular pairs of islands at each vertex, vertices of types 5 and 6 (Fig. \ref{fig:vertexfiglabels}) were energetically favoured and the ground state showed antiferromagnetic order \cite{Moller2006,Morgan2011,Levis2013}. For this reason, it has been suggested that square-lattice ASI systems may be approximately described using the F-model \cite{NisoliColloquium2013}. It is interesting therefore to ask whether the F-model phenomenology presented in this article is relevant to these systems.

In general, the behaviour of square-lattice ASI systems is determined by the relative strengths of the ferromagnetic interactions between nearest (perpendicular) neighbour $J_1<0$ and between next-nearest (opposite) neighbours $J_2<0$ at each vertex. Typically~\cite{Moller2006,Perrin2016} the ratio $J_2/J_1$ takes values between 0.5 and 0.7, but recent work has demonstrated significant scope to tune $J_2/J_1$ beyond this interval by introducing height modulation~\cite{Perrin2016}, interaction modifiers~\cite{Ostman2018} or by introducing perforations at the vertices in connected ASI structures~\cite{Schanilec2019}.

Interestingly the limits $J_{2}/J_{1} \to 1^{-}$ and $J_{2}/J_{1} \to 0$ are exactly solvable, corresponding respectively to the ``pure’’ F-model and a ``modified’’ F-model~\cite{Wu1969,LiebWu1972} with a second-order transition at $T_{0}^{\rm (Mod)}=\epsilon /(2 \ln (\sqrt{2}+1))$. The ability to tune the ratio $J_{2}/J_{1}$  thus offers the enticing prospect of using experimental ASI systems to interpolate between the exactly solved limits and elucidate how the second-order transition evolves into the infinite-order transition of the F-model as the ``pure’’ limit is approached. These exactly-solvable limits for artificial square ice are discussed further in Appendix \ref{AppendixASI}.

Artificial spin ice systems further offer an excellent experimental system in which to probe the effect of different boundary conditions on vertex models which, unusually, are expected to impact the free energy of these systems even in the thermodynamic limit and can lead to exotic phenomena such as arctic curves \cite{Korepin2000, ZinnJustin2000, ZinnJustin2002, Bogolyubov2003, Colomo2010, Cugliandolo2014}. Fixed boundary conditions could readily be imposed in experimental ASI systems simply by elongating the ferromagnetic islands that make up the boundary so that their blocking temperature is higher than for islands in the bulk, while recently developed techniques to reverse the magnetization of individual islands \cite{Gartside2018,Gartside2020} could be used to impose arbitrary boundary conditions.

A final mention should be made of ASI systems constructed on the unconventional ``Shakti’’ lattice \cite{Morrison2013}, for which the ground state is dual to a thermal state of the F-model \cite{Chern2013,Gilbert2014} at some fictitious temperature. Other ASI systems on unconventional lattices have been fruitfully described in terms of topological invariants \cite{Lao2018,Zhang2020} and it would be worth exploring whether topological sector fluctuations in such systems may similarly be linked directly to experimentally accessible quantities.

F-model physics may thus unexpectedly pop up in seemingly unrelated artificial spin systems, while square-lattice ASI might be used to explore how the modified F-model approaches the pure F-model by tuning the interactions between nearest and next-nearest neighbours to alter the energies associates with ice-rule defects. In the latter case it will be particularly interesting to observe at what point a description in terms of topological sectors becomes useful, clarifying how meandering magnetic monopoles observed previously \cite{Morley2019} may facilitate fluctuations between topological sectors. In this regard it will be useful to determine the system parameters over which the universal jump and breakdown of the Fisher relation discussed above appear.

\section{Conclusions}

In conclusion, antiferrolectrics and their analogues are typically described by Landau theory and expected to have conventional phase transitions and wavelike excitations. Ice rule systems, on the other hand, provide examples of phases and transitions that go beyond this paradigm. Most notably, Powell has shown~\cite{Powell2011} how they provide examples of transitions that are well described as the Higgs transitions of an emergent gauge theory. In general, the properties of ice model systems may be described in terms of their topological defects: these control, for example, not only the equilibrium behaviour, but also the out-of-equilibrium behavior, for example the slow time evolution of the system following thermal quenches into the ordered or disordered ground state~\cite{Levis_EPL_2012}. 

In this paper, we have focused on properties arising from one type of topological defect, corresponding to ferroelectric or ferromagnetic windings. These form topological sectors and their effects are emphasised by the particular constraints of the F-model, which exclude `monopole' defects. By extending existing theories of the F-model, we have established the experimental signatures of these winding defects, identified their novel consequences (such as the universal jump in their fluctuations) and clarified some aspects of theory. 

It is clear from our non-exhaustive survey of experimental systems that the topological sector fluctuations of the F-model are highly relevant to many experiments. Thus, realistic ice rule systems can be found that should reveal the topological properties of the F-model to an excellent approximation. In the future it will be very interesting to see to what extent the classic theories of Lieb and Baxter, as we have articulated them, can be accurately be realised in experiment. 

\acknowledgements{It is a pleasure to thank Callum Gray for fruitful discussions, and Laura Foini and Demian Levis for helpful correspondence regarding the cavity method.

We further wish to thank Cristiano Nisoli for useful discussions regarding his Faraday loop representation of the F-model.

Finally, we thank the Leverhulme Trust for their support through grant RPG-2016-391.}

\appendix

\section{Exactly-Solvable Limits for Artificial Square Ice}\label{AppendixASI}

As noted in Section III, contact may be made with two exactly-solvable models by tuning the interactions between the nanomagnetic elements of artificial square ice systems. In general, the dipolar interactions between the bistable moments $\sigma_{\rm i}=\pm1$ lead to a Hamiltonian of the form

\begin{equation}
    \mathcal{H} = \sum_{<\rm i,\rm j>} J_{\rm ij}\sigma_{\rm i}\sigma_{\rm j}
\end{equation}
where $J_{\rm ij}$ encodes the strength of the interaction between pairs of moments $\sigma_{\rm i}$ and $\sigma_{\rm j}$. Since the range of the dipolar interaction is infinite the sum is in principle over all pairs of moments, but M\"{o}ller and Moessner~\cite{Moller2006a} argue that in suitably fabricated systems the Hamiltonian may be truncated to include only nearest-neighbour (perpendicular) and next-nearest-neighbour (parallel) pairs with interaction strengths $J_{1}$ and $J_{2}$, respectively. For individual vertices with $J_{1}<J_{2}<0$ this gives an energy hierarchy similar to the F-model with energy gap $\epsilon=8(J_{2}-J_{1})$, supplemented by 3/1 defects and 4/0 defects with energies

\begin{eqnarray}
    \frac{\epsilon_{\rm 31}}{\epsilon} &= \frac{(J_{2}/J_{1}-2)}{2(J_{2}/J_{1}-1)}\\
    \frac{\epsilon_{\rm 40}}{\epsilon} &= \frac{-2J_{1}}{(J_{2}-J_{1})}
\end{eqnarray}

In the limit where nearest-neighbour interactions dominate $J_{2}/J_{1}\to0$ we thus obtain the standard F-model energy hierarchy supplemented with $\epsilon_{\rm 31} =\epsilon$ and $\epsilon_{\rm 40} = 2\epsilon$. Since this ``modified F-model’’  \cite{Wu1969a,LiebWu1972a} is simply the standard square-lattice Ising model, it has an exact solution with a second-order transition at $T_{0}^{\rm (Mod)}=\epsilon /(2 \ln (\sqrt{2}+1))$. As for the ``pure’’ F-model there is a maximum in the zero-field susceptibility which here occurs around $1.537\; T_{0}^{\rm (Mod)}$ (Refs. \cite{Sykes1962a,Orrick2001a}), though unlike the pure model the susceptibility does not vanish below $ T_{0}^{\rm (Mod)}$ since broken loops that incorporate defects may carry a net polarization. Introducing significant populations of ice-rule defects thus breaks the mapping between topological sectors and net polarization, so that we cannot expect artificial square ice systems in which defects have similar energies to ice-rule vertices to be well-described by the pure F-model despite their shared ground state.

In order to bring square-lattice ASI systems closer to the pure F-model the energy associated with ice-rule defects must therefore be increased relative to $\epsilon$. This may be achieved by bringing $J_{1}$ and $J_{2}$ as close together as possible ($J_{2}/J_{1} \to 1^{-}$) while keeping $J_{1}<J_{2}<0$. In this case we still have $\epsilon - 8(J_{2} - J_{1})$ while $\epsilon_{\rm 31}/\epsilon \approx - J_{1}/(2(J_{2} - J_{1}))$ and $\epsilon_{\rm 40}/\epsilon = - 2 J_{1}/(J_{2} - J_{1})$, so that the energy cost of introducing ice-rule defects increases in proportion to $(1 - J_{2}/J_{1})^{- 1}$ until $J_{1}=J_{2}$ where the ice model is recovered. The latter case has already been achieved experimentally in impressive work by Perrin et al.~\cite{Perrin2016a} that introduced a height offset between the vertical and horizontal sublattices of square ice to equalize $J_{1}$ and $J_{2}$, a result recently replicated by Farhan et al.~\cite{Farhan2019a} and explored in a different system by Lehmann et al. \cite{Lehmann2020a}. If the interactions were tuned to $J_{2}/J_{1}=0.99$ the energy cost of introducing defects would be $\epsilon_{\rm 31}\approx50\epsilon$ and $\epsilon_{\rm 40}=200\epsilon$, so that we can expect a substantial temperature window in which the number of defects is small and the pure F-model is a good approximation.

\end{document}